\documentclass{svmult}

\usepackage{graphicx}    

\usepackage{amssymb}                        

\def\beq{\begin{equation}}
\def\eeq{\end{equation}}

\def\bea{\begin{eqnarray}}
\def\eea{\end{eqnarray}}
\def\ba{\begin{array}}                  
\def\ea{\end{array}}

\def\a{\alpha}

\def\L{\Lambda}

\begin{document}

\title*{Noncommutative Gauge Theory on the $q$-deformed Euclidean Plane}

\titlerunning{Noncommutative Gauge Theory on the $q$-deformed Euclidean Plane} 
\author{Frank Meyer\inst{1,2} \and
Harold Steinacker \inst{2}}
\institute{Max-Planck-Institute for Physics (Werner-Heisenberg-Institut), 
F\"ohringer Ring 6, D-80805 M\"unchen, Germany
\texttt{meyerf@mppmu.mpg.de}
\and University of Munich, Theresienstr. 37, D-80333 M\"unchen, Germany 
\texttt{hsteinac@theorie.physik.uni-muenchen.de}  }

%
%
\maketitle

\section{Introduction} \label{sec:intro}

The physical nature of space at very short distances is still not known.
Already Heisenberg proposed in a letter to Peierls \cite{letter_to_Peierls}
that spacetime is quantized below some scale, suggesting that 
this could help to resolve
the problem of infinities in quantum field theories. 
With this motivation, 
there has been a lot of work and progress in the 
formulation of quantum field theory on quantized or noncommutative spaces.
Noncommutativity is implemented by replacing a differentiable 
space-time manifold by an algebra of noncommutative coordinates 
\beq \label{eq:noncomm}
[x^i,x^j] = \theta^{ij}(x) \neq 0 \,.
\eeq
The simplest case is the so-called canonical quantum plane 
$\mathbb R^n_\theta$, 
where $\theta^{ij}$ is a constant tensor independent of $x$.
This is the space which is 
usually considered in the literature \cite{Douglas}.
However, most of the rotational symmetry is lost on $\mathbb R^n_\theta$. 
On the other hand,
there exist quantum spaces which admit a generalized notion of symmetry,
being covariant under a quantum group. 
Not much is known about field theory on this type of spaces.

One of the simplest spaces with quantum group symmetry is 
the Euclidean quantum plane $\mathbb{R}_q^2$. It 
is covariant with respect to the 
$q$-deformed two-dimensional Euclidean group $E_q(2)$. 
We report here on our work \cite{paper}, proposing a formulation of gauge theory 
based on the natural algebraic structures on this spaces and 
using a suitable star product.

\section{The $E_q(2)$-Symmetric Plane}
The $E_q(2)$-Symmetric Plane 
is generated by the complex
coordinates $z,\bar{z}$ with the commutation relation
\beq\label{eq:comEq(2)}
z\bar{z}=q^2\bar{z}z \,.
\eeq
We consider formal power series in these variables $z,\bar{z}$
as functions on this space, 
\beq
\mathbb{R}_q^2:=\mathbb {R} \langle \langle z,\bar{z} \rangle
\rangle/(z\bar{z}-q^{2}\bar{z}z) \,. 
\eeq
Notice that the simple commutation relation (\ref{eq:comEq(2)}) 
are inconsistent with the
usual formulas for differentiation and integration. 
We should therefore first discuss the appropriate 
differential calculus and invariant integration.
Finally, to get physical predictions i.e. real numbers from 
 the abstract algebra, we also need either
 a representation of the noncommutative algebra, 
or a realization of the algebra using a star product.

\subsection{Covariant Differential Calculus}

It is natural to require that there exist deformed spaces of $k$-forms 
$\Omega^k_q$ which are covariant with respect to $E_q(2)$, 
and that the exterior differential 
$d: \Omega^k_q \rightarrow \Omega^{k+1}_q$ 
satisfies the
usual Leibniz-rule as well as $d^2=0$. One can show that
there exists a unique 
covariant differential calculus with these properties \cite{Chaichian}:
\begin{equation}
\label{eq: comm-rel dz,dzbar with z,zbar}
\begin{array}{cccccccc}
zdz & = & q^{-2}dzz, &  &  & \overline{z}dz & = & q^{-2}dz\overline{z}\\
zd\overline{z} & = & q^{2}d\overline{z}z, &  &  & \overline{z}d\overline{z} & = & q^{2}d\overline{z}\, \overline{z} \\
\end{array}
\end{equation}
The following result is particularly useful for the construction of 
gauge field theories on $\mathbb{R}_q^2$:
\begin{lemma}
Consider the one-forms
\beq
\theta \equiv \theta^{z} := z^{-1}\overline{z}dz \quad , \quad \overline{\theta } \equiv \theta^{\overline{z}} := d\overline{z}z\overline{z}^{-1} \,
\eeq
and define 
\beq
\Theta :=\theta ^{i}\lambda _{i}, \quad \mbox{where} \quad
\lambda _{z} :=\frac{1}{1-q^{-2}}\overline{z}^{-1}\, , \, \lambda _{\overline{z}} := -\frac{1}{1-q^{-2}}z^{-1}.
\eeq 
Then for all functions \( f\in \mathbb {R}_{q}^{2} \) and one-forms $\alpha$,
the following holds:
\bea
[\theta ,f] &=& [\overline{\theta },f]=0,   \\
df &=& [\Theta ,f]=[\lambda _{i},f]\theta ^{i}
\label{Thetacomm} \\
d\alpha &=&\{\Theta ,\alpha \}
\label{Thetacomm-1}
\eea
denoting with $\{\cdot,\cdot\}$ the anti-commutator.
\end{lemma}

\subsection{Invariant Integral}
In order to define an invariant action, we need an integral on 
\( \mathbb {R}_{q}^{2} \) which is invariant
under quantum group transformations. 
This means that
\begin{equation}
\label{eq: invariance condition for the q-integral}
\int ^{q}f(z,\overline{z})\triangleleft X=\varepsilon (X)\int ^{q}f(z,\overline{z})
\end{equation}
for all \( f \in \mathbb{R}^2_q\) and \( X\in U_{q}(e(2)) \). 
Here \( U_{q}(e(2)) \) is the $q$-deformed universal enveloping Lie algebra 
of the two-dimensional Euclidean
group, $\triangleleft$ denotes the right action of \( U_{q}(e(2)) \) on \(\mathbb{R}^2_q\) and 
$\varepsilon (X)$ is the counit. Now any function 
in \(\mathbb{R}^2_q\)  can be decomposed as 
\(f(z,\overline{z})=\sum _{m\in \mathbb {Z}}z^{m}f_{m}(z\overline{z}). \)
It can be shown that (\ref{eq: invariance condition for the q-integral}) 
is satisfied for the following discrete
quantum traces \cite{Koelink}
\begin{equation} \label{inv_int}
\int^{q,(r_0)}f(z,\overline{z}):=r^{2}_{0}(q^{2}-1)\sum ^{\infty
}_{k=-\infty }q^{2k}f_0(q^{2k}r_{0}^{2}),
\end{equation}
where $r_0 \in \mathbb{R}$ labels the irreducible representations of \(\mathbb{R}^2_q\). 
The most general invariant integral is given by
superpositions of these integrals,
\begin{equation}
\label{de: Eq(2) invariant q-integral 1}
\int^{q} f(z,\overline{z}) = \int_1^q dr_0 \mu(r_0) \int^{q,(r_0)}f_0(z\overline{z}) 
\end{equation}
with arbitrary  "weight" function $\mu(r) >0$. 
It is quite  remarkable and 
useful that for the special choice $\mu(r_0)=\frac 1{r_{0}(q^{2}-1)}$, 
one recovers 
the usual  Riemannian integral, which is therefore
also invariant under  \( U_{q}(e(2)) \) \cite{paper}. 

\section{Star Product Approach}
\subsection{The Star Product}
The noncommutative algebra $\mathbb{R}^2_q$ can be realized
on the algebra of commutative functions on
$\mathbb{R}^2$ using a new, noncommutative product, 
called star product. Let us denote the commutative variables on 
$\mathbb{R}^2$ by greek letters $\zeta, \bar{\zeta}$ to distinguish
them from the generators $z,\bar{z}$, and let $q=:e^h$. Then
a hermitian star product for $\mathbb{R}^2_q$ is given by
\begin{equation}
\label{de: q-symm star product}
f\star g:=\mu \circ e^{h(\zeta \partial_{\zeta}\otimes 
\overline{\zeta}\partial_{\overline{\zeta}}-\overline{\zeta}\partial_{\overline{\zeta}}\otimes \zeta\partial
_{\zeta})}(f\otimes g)=fg+h\zeta \overline{\zeta}(\partial_{\zeta} f \partial_{\overline{\zeta}} g -
\partial_{\overline{\zeta}} f \partial_{\zeta} g) +
\mathcal{O}(h^2) \, .
\end{equation}
\subsection{Noncommutative Gauge Transformations}
The formalism of covariant coordinates was established in 
\cite{Madore} for an arbitrary Poisson structure. 
This leads to problems in the semi-classical limit\footnote{To obtain in the classical limit the classical gauge
field $a_i$ we have to invert $\theta^{ij}$. This is only well-defined if $\theta$ is invertible, and even then
it spoils the covariant transformation property whenever $\theta$ is not constant. To maintain covariance one
has to "invert $\theta$ covariantly" as done in \cite{Diploma}, leading to complicated expressions.}.
Therefore we propose the 
following approach, taking advantage of the
frame $\theta, \bar{\theta}$ which commutes with all functions 
and the generator $\Theta$ of the exterior differential. 
We define infinitesimal gauge transformations of a matter 
field as 
\beq
\delta \psi = i \Lambda \star \psi , \qquad \delta \zeta^i=0 \,.
\eeq
Let us introduce the "covariant derivative" (or covariant one-form) as 
\beq
D:=\Theta -i A \,.
\eeq  
Then requiring that $D\psi$ transforms covariantly, i.e.
\beq
\delta D\star \psi \stackrel{!}{=}i\Lambda \star D \star \psi
\eeq
leads to the following gauge transformation property for the gauge field $A$:
\beq
\delta A=[\Theta \stackrel{\star}{,}\Lambda]+i[\Lambda \stackrel{\star}{,}A]=d\Lambda +i[\Lambda
\stackrel{\star}{,}A] \,.
\eeq
We define the field strength as the two-form 
\beq
F:=D \wedge_q D
\eeq
where $\wedge_q$ the star-wedge \cite{paper}. Then
\beq
\delta F = i[\Lambda \stackrel{\star}{,}F].
\eeq
As a two-form, the field strength can be written as 
$F=f\theta \wedge_q \bar{\theta}$. Since the frame $\theta,
\bar{\theta}$ commutes with functions, $f$ transforms covariantly as well:
\beq \label{gaugetrafo_f}
\delta f= i[\Lambda\stackrel{\star}{,}f] \,.
\eeq
We note that all transformations have the correct 
classical limit as $q \rightarrow 1$.

\subsection{Seiberg-Witten Map}
The Seiberg-Witten map \cite{SW} allows 
to express the noncommutative gauge fields in terms of
the commutative ones. Hence the noncommutative theory can be 
interpreted as a deformation of the
commutative theory. Its physical predictions can be explicitly obtained
by expanding in the deformation parameter $h$, and the commutative theory is
reproduced in the limit $h\rightarrow0$.
The Seiberg-Witten map is based on the following requirement:
\begin{itemize}
\item The \emph{consistency condition:}
\begin{equation}
\label{eq:consistency}
\begin{array}{cccc}
 & (\delta _{\alpha }\delta _{\beta }-\delta _{\beta }\delta _{\alpha })\Psi  & = & \delta _{-i[\alpha ,\beta ]}\Psi \\
\Leftrightarrow  & i\delta _{\alpha }\Lambda _{\beta }-i\delta _{\beta }\Lambda _{\alpha }+[\Lambda _{\alpha }\stackrel{\star}{,}\Lambda _{\beta }] & = & i\Lambda _{-i[\alpha ,\beta ]},
\end{array}
\end{equation}
\item Noncommutative gauge transformations are related to commutative ones:
\begin{eqnarray}
A_{i}[a_{i}]+\delta_{\L} A_{i}[a_{i}] & = & A_{i}[a_{i}+\delta_{\a} a_{i}]\label{eq:SWequ} \\
\Psi [\psi ,a_{i}]+\delta_{\L} \Psi [\psi ,a_{i}] & = & \Psi [\psi +\delta_{\a} \psi ,a_{i}+\delta_{\a} a_{i}]\, .\nonumber 
\end{eqnarray}
\end{itemize}
Solving these conditions in our case, we obtain  the following
expression for the field strength expanded in powers of $h$:
\begin{eqnarray}
f &=& F^0_{12}+h\big \{F^0_{12}+
\theta^{12}(F^{0}_{12}F^0_{12}-a_{\zeta}\partial_{\overline{\zeta}}F^0_{12}+a_{\overline{\zeta}}\partial_{\zeta}F^0_{12})
+\partial_{\zeta}\theta^{12}(a_{\zeta}\partial_{\overline{\zeta}}a_{\overline{\zeta}}
 \label{SW_f} \\ 
& &
+a_{\overline{\zeta}}\partial_{\overline{\zeta}}a_{\zeta}+2a_{\overline{\zeta}}\partial_{\zeta}a_{\overline{\zeta}})+\partial_{\overline{\zeta}}\theta^{12}(a_{\zeta}\partial_{\zeta}a_{\overline{\zeta}}+a_{\overline{\zeta}}\partial_{\zeta}a_{\zeta}+2a_{\zeta}\partial_{\overline{\zeta}}a_{\zeta})
\big \}+\mathcal{O}(h^2) \, ,
\nonumber
\end{eqnarray}
where $a_{\zeta_i}$ is the classical gauge field and $F^0_{ij}=\partial_{\zeta_i}a_j-\partial_{\zeta_j}a_i$ is
the classical field strength. 

\subsection{The Action}

To define a gauge-invariant action, 
we need an integral which is cyclic with respect to the star product,
since the field strength transforms in
the adjoint. The invariant integrals (\ref{inv_int}) 
do not have this property, which can be restored by introducing
a measure function $\mu$. It can be shown that for the measure function
$\mu:=\frac{1}{\zeta\bar{\zeta}}$, we can even drop the 
star under the integral: 
\beq
\int d\zeta d\bar{\zeta} \mu f\star g = \int d\zeta d\bar{\zeta} \mu fg = \int d\zeta d\bar{\zeta} \mu g\star f 
\eeq
Putting all this together, we can now define a gauge invariant action:
\beq
S:=\frac{1}{2}\int d\zeta d\bar{\zeta} \mu f\star f \,.
\eeq
Gauge invariance is guaranteed because of (\ref{gaugetrafo_f}). Moreover, the classical action for abelian
gauge field theory is reproduced in the classical limit $h\rightarrow 0$ because of (\ref{SW_f}).

Another possibility is to replace 
the measure function $\mu$ by a scalar ``Higgs'' 
field $\varphi$, which transforms
such that it restores the gauge invariance of the action.
One can in fact find a suitable potential for $\phi$ which admits 
a solution $\langle \phi \rangle = \mu = \frac{1}{\zeta\bar{\zeta}}$, 
leading to a $E_q(2)$-invariant action through spontaneous symmetry breaking.

%

\end{document}